# An Analytical Model for Service Profile Based Service Quality of an Institutional eLibrary

Ash Mohammad Abbas
Department of Computer Engineering
Aligarh Muslim University
Aligarh 202002, India
Email: am.abbas.ce@amu.ac.in

*Abstract*—Devising a scheme for evaluating the service quality of an institutional electronic library is a difficult and challenging task. The challenge comes from the fact that the services provided by an institutional electronic library depend upon the contents requested by the users and the contents housed by the library. Different types of users might be interested in different types of contents. In this paper, we propose a technique for evaluating the service quality of an institutional electronic library. Our scheme is based on the service profiles of contents requested by the users at the server side which is hosted at the library. Further, we propose models to analyze the service quality of an electronic library. For analyzing the service quality, we present two analytical models. The first one is based on the number of days by which the item to be served by the library is delayed and the penalty points per day for the duration for which the item is delayed. The second model is based on the credits earned by the library if the item is served in a timely fashion, and the penalties, thereof, if the item is delayed. These models may help in evaluating the service quality of an electronic library and taking the corrective measures to improve it.

## I. INTRODUCTION

The proliferation of Internet, specifically, the World Wide Web (WWW) has made a tremendous impact on the society in terms of what we need, how we act, and our habits. For example, instead of going to a physical library, we now wish to retrieve the contents at our desktops, laptops, or on mobile devices. We wish to have ubiquitous access to the Internet irrespective of whether we are moving or even traveling from one part to the other part of the world. If we want to read a book, we want that the book should be readily available at the device we are using at the moment. This has led to a concept of electronic library (eLibrary), where the contents can be downloaded to the device by simply clicking the mouse or just pressing some buttons.

The physical libraries in todays world cannot survive if they do not provide the contents electronically. Therefore, in addition to housing physical books, a physical library should also possess electronic books (ebooks), videos, lecture slides, thesis, reports, journals in electronic form (ejournals), etc. The library has to be connected to the Internet so that internal users (or users from inside the institute) and external users, including those from different parts of the world, may have an access to the resources and contents of the library. In other words, a modern library has to act as a content provider, rather than traditionally providing only books, journals, reports, and thesis, all in physical form.

Many of the researchers have focused on what types of contents should be provided by an eLibrary. In [1], the impact of query correlation and query semantics on the information retrieved from online digital libraries is described. A set of guidelines and criteria for selecting the electronic resources is available at [2]. The way, the libraries use the print and electronic resources is discussed by Shorten (2006). The design of a middleware for building adaptive systems, called DISSelect-based Adaptive System (DISSAS), has been described in [3], which can be used to enable adaptation in web-based applications and legacy information systems. In [6], an e-content selection method using multiple criteria analysis in web-based personalized learning environments is described. The work in [9] presents a timely and keyword-based dynamic content selection for public displays. In [4], a user driven content selection scheme for digitization of Ebooks-on-Demand (EoD) networks is presented.

A conceptual model in the form of a questionnaire called *ServQual* to evaluate the service quality of a system was presented in [8]. The ServQual was, originally, comprising of ten aspects of service quality, namely, *reliability*, *responsiveness*, *competence*, *access*, *courtesy*, *communication*, *credibility*, *security*, *understanding the customer*, and *tangibles*. It was aimed to measure the gap between customer expectations and experience. Later, in [12], the model was refined to contain only five attributes: **r**eliability, **a**ssurance, **t**angibles, **e**mpathy, and **r**esponsiveness; and was renamed as RATER, which is an acronym for the set of attributes it contains. Since their introductions, the ServQual and RATER are used for assessing the service quality in different fields such as health care, business, financial, marketing, etc. The questionnaires of ServQual or RATER can also be used for determining the service quality of a library. However, these tools ServQual and RATER are generalized to evaluate the service quality of any system, and not specifically of a library system. A specific tool for evaluating the service quality of a library, called "LibQUAL+®" is described in [13]. Item sampling in service quality assessment surveys to improve the response rate and reduce the burden on the respondent using a tool called "LibQUAL+® Lite" is studied in [14]. A brief comparison of these tools is presented in Table I. A common feature of these tools is that the assessment of service quality of an underlying system is based on the outcomes of the surveys.

In [7], an assessment of the service quality of Thammasat University Library System is studied using a modified version of the questionnaire of ServQual. The authors, therein, use a concept of *zone of tolerance*. They conducted a survey on different classes of users such as undergraduate students, graduate students, faculty members, and researchers. They consider organizational, access, and personal effects on the service quality of the library by manually counting and categorizing the problems that users face in each of the three categories. The effect of individual differences such as gender and status of a private university library namely, Independent University Bangladesh Library, is investigated in [10]. The authors, therein, carried out a survey using a modified version of the questionnaire of ServQual, and their findings suggest that scores of different classes of users may differ based on their gender and status. An assessment of service quality at central library of Management and Planning Organization (MPO), Iran, is carried out in [5]. Their study focuses on the factors related to the library environment, information dissemination, and library personnel. They suggest that a focus on the training and development of library staff may help in providing better services.

However, most the evaluations of service quality of a library (e.g. [7], [10], [5]) are carried out for physical libraries. Not much work is available in the literature for electronic libraries. As we mentioned earlier, there is a paradigm shift from physical libraries to electronic libraries. However, we expect that physical libraries shall continue to exist, at least in the near future, due to many reasons. For example, one reason can be that not all physical contents can be converted to their truly electronic form, and the kind of entertainment by visiting physical libraries and looking at physical objects cannot be attained by simply watching their electronic form. The only thing is that many of the users would like to get the electronic contents, however, some other users may prefer to visit the library physically depending upon the availability of time and depending upon tightness their schedules. Libraries should serve both classes of users in the best possible manner, as a result, their comes a concept of the service quality of a library, be it electronic or physical, which is the theme of this paper.

In this paper, we propose analytical models for evaluating the service quality of an institutional electronic library. Our models are based on the service profiles of contents requested by the users. Specifically, we propose two models for the service quality. The first model is based on the duration by which an item to be served by the library is delayed and the penalties thereof. The second model is based on the credits if the items requested by users are served in a timely fashion and the penalties if the items to be served by the library are delayed. Our model is not based on any servey, however, it is based on how the request is handled by the library. In other words, our model is based on the requests served by the library and the service quality is evaluated on the basis of the requests served.

The rest of this paper is organized as follows. In section

TABLE I
A COMPARISON OF THE TOOLS FOR SERVICE QUALITY ASSESSMENT

| Tools | Basis | Features | Comments |
|---|---|---|---|
| ServQual [8] | Survey | 10 aspects of service quality | Generalized to any system |
| RATER [12] | Survey | 5 aspects of service quality | Generalized to any system |
| LibQUAL+® [13] | Survey | Item Sampling | Specific to Library |

II, we describe the notion of service profiles for the requests received by the library for specific contents. In section III, we present analytical models for service profile based service quality. In section IV, we present results and discussion. The last section is for conclusion.

## II. SERVICE PROFILES

In this section, we propose a service profile based scheme for evaluating the service quality provided by the eLibrary.

A library maintains the service profile about the services provided to users and also to different categories of users so as to improve its services in the future. The service profile contains the information about the services provided by the eLibrary. Specifically, a service profile of an eLibrary contains the following open ended set of attributes.

<*RequestID, RequestTime, UserID, ContentID, ContentType, ContentHits, ContentAvailStatus, ContentDeliveryTime, ArrangementStatus, NotificationStatus, NotificationTime, UserAcceptance, ReasonsNotDelivered, ExcessDelay*>.

The attribute *RequestID* is an identifier for the request generated by an end-user. The attribute *RequestTime* represents the time of the reception of the user request by the eLibrary. The attribute *UserID* is an identifier of the user who generated the request. The attribute *ContentID* is an identifier for the content requested by the user, and the attribute *ContentType* represents the type of the content the user has requested such as physical book, ebook, video, ppt slides, journal, tutorials, reports, thesis, etc. The attribute *ContentHits* represnts the number of user requests received for a specific content within a specified observation time. The attribute *ContentAvailabilityStatus* tells whether the content is available or not available. If the content is available, then the attribute *ContentDeliveryTime* tells the time when the content is delivered to the user. Otherwise, the attribute *ArrangementStatus* tells whether the content will be arranged/procured by the eLibrary or not. If the arrangement/procurement of the content will be carried out by the eLibrary, then the expected time the arrangement/procurement is going to incur. The attribute *NotificationStatus* tells whether the notification is sent to the user or not, and the attribute *NotificationTime* represents the time when the notification was sent to the user informing him about the arrangement/procurement. The attribute *UserAcceptance* represents whether the user agrees to the time taken by the eLibrary in arrangement/procurement of the content. If the content is not delivered at all, then the reasons are recorded for not delivering the content to the user in the *ReasonsNotDelivered* field. The attribute *ExcessDelay* represents the delay in excess to what has been agreed between the user and the eLibrary.

In what follows, we present models for analyzing the service profiles based quality of service provided by an eLibrary.

## III. ANALYSIS OF SERVICE PROFILES

For analyzing the service quality of an eLibrary, we present two models: (i) delay based service quality model, and (ii) credit based service quality model. We describe each of them as follows.

### A. Delay Based Service Quality Model

This model is based on the absolute delays (say, in days) between the day on which the request was made by the user or the item was due to be delivered, and the day on which request was actually serviced.

Let $\tau$ be the *ExcessDelay*, in number of days, incurred after the expiry of the expected time of delivery as notified by the eLibrary to the user, and p be the penalty points per day assigned by the library itself, with the viewpoint to evaluate the service quality provided by the eLibrary. Let $\phi(p, \tau)$ be the service quality of the eLibrary with parameters $p$ and $\Delta$ and let it be given by the following expression.

$$\phi(p,\tau) = 1 + pe^{-p\tau} \quad (1)$$

where, $p \geq 0$, $\tau \geq 0$. In this model, the maximum value of the service quality is,

$$\phi_{\max} = 1 + p. \quad (2)$$

The maximum value of service quality occurs when the parameter $\tau = 0$. The minimum value of the service quality is $\phi_{\min} = 1$ and occurs at $p = 0$. The service quality is calculated for all requests the eLibrary receives and then the average value of the service quality can be determined by taking the average over all requests considered, which is expressed as follows.

$$\bar{\phi} = \frac{\sum_{i=1}^{n} \phi_i}{n}. \quad (3)$$

To discuss how the service quality varies with the variations in the parameters $p$ and $\tau$, we need to compute the partial derivatives of the service quality with respect to these parameters. We compute the partial derivatives of the service quality with respect to the parameters $p$ and $\tau$ in the following lemma.

*Lemma 1:* The partial derivative of the service quality with respect to $\tau$ is as follows.

$$\frac{\partial \phi}{\partial p} = (1 - p^2)e^{-p\tau} \quad (4)$$

and,

$$\frac{\partial \phi}{\partial \tau} = -p^2 e^{-p\tau}. \quad (5)$$

*Proof:* The partial derivative of the service quality with respect to $p$ is as follows.

$$\begin{aligned}\frac{\partial \phi}{\partial p} &= pe^{p\tau}(-p) + e^{-p\tau} \\ &= -p^2 e^{-p\tau} \\ &= (1 - p^2)e^{-p\tau}.\end{aligned} \quad (6)$$

Similarly, the partial derivative of the service quality with respect to the excess delay $\tau$ is as follows.

$$\begin{aligned}\frac{\partial \phi}{\partial \tau} &= pe^{p\tau}(-p) \\ &= -p^2 e^{-p\tau}.\end{aligned} \quad (7)$$

■

Let the variation in the service quality with respect to $p$ be $\Delta\phi_p$, and the variations in the service quality with respect to the variations in $\tau$ be $\Delta\phi_\tau$, then the overall variations in the service quality is as follows.

$$\Delta\phi = \frac{\partial \phi}{\partial p}\Delta\phi_p + \frac{\partial \phi}{\partial \tau}\Delta\phi_\tau. \quad (8)$$

In the above model, the penalty for an item delivered late has been incorporated and the effective penalty varies with the duration by which the item is delivered late. However, the above model does not take into account any credits for the items delivered in time.

We now present a credit based service quality model.

### B. Credit Based Service Quality Model

Let $H$ be the number of requests served in order (i.e. on the same day or on or before the day mutually agreed between the eLibrary and the user). Let $L$ be the number of requests served late (i.e. after the mutually agreed day or time between user and the eLibrary). Let there be $c$ number of credits assigned for each request served in time. If the time of service of the request is delayed, then a penalty $p$ is imposed on to the eLibray. Note that the total number of requests received by the eLibrary is the summation of the number of requests served in time and the number of requests served late, i.e. $H + L$. We now define the service quality as follows.

$$\phi = \frac{cH - pL}{(c+p)(H+L)} \quad (9)$$

where, $-1 < \phi < 1$. If $c = p = q$, then the expression of the service quality becomes as follows.

$$\phi = \frac{H - L}{2(H + L)} \quad (10)$$

Note that when $c = p = q$, and $H = L$, we have $\phi = 0$; and if $L = 0$, $\phi = \frac{1}{2}$; similarly, if $H = 0$, $\phi = -\frac{1}{2}$. We can now say that for $c = p$, $-\frac{1}{2} \leq \phi \leq \frac{1}{2}$.

To discuss how the service quality varies with the number of requests served in a timely fashion and the number of requests served late, we need to compute the partial derivatives of the service quality with respect to $H$ and $L$, respectively. We prove the following lemma about the derivatives of the service quality with respect to the number of requests served late as well as the number of requests served in a timely fashion.

*Lemma 2:* The partial derivatives of the service quality with respect to the number of requests served late as well as with respect to the number of requests served in a timely fashion are given by,

$$\frac{\partial \phi}{\partial H} = \frac{L}{(H+L)^2} \quad (11)$$

and
$$\frac{\partial \phi}{\partial L} = -\frac{H}{(H+L)^2}. \quad (12)$$

*Proof:* Using the law of division, the partial derivative of the service quality, as defined by (9), with respect to $H$, is as follows.

$$\begin{aligned}
\frac{\partial \phi}{\partial H} &= \frac{(c+p)(H+L).c - (cH-pL).(c+p)}{\{(c+p)(H+L)\}^2} \\
&= \frac{c(c+p)H + c(c+p)L - (c+p)cH + (c+p)pL}{\{(c+p)(H+L)\}^2} \\
&= \frac{(c+p)^2 L}{\{(c+p)(H+L)\}^2} \\
&= \frac{L}{(H+L)^2}. \quad (13)
\end{aligned}$$

Similarly, the partial derivative of the service quality with respect to $L$ is given by,

$$\begin{aligned}
\frac{\partial \phi}{\partial L} &= \frac{(c+p)(H+L).(-p) - (cH-pL).(c+p)}{\{(c+p)(H+L)\}^2} \\
&= \frac{-p(c+p)H - p(c+p)L - c(c+p)H + (c+p)pL}{\{(c+p)(H+L)\}^2} \\
&= \frac{-(c+p)^2 H}{\{(c+p)(H+L)\}^2} \\
&= -\frac{H}{(H+L)^2}. \quad (14)
\end{aligned}$$

∎

From the expressions (11) and (12), it is clear that partial derivatives do not depend on the number of credits, $c$, or penalty points, $p$, and depend only on how many requests were served in a timely fashion and how many requests were served late.

One can utilize the partial derivatives to compute the variation in the service quality. Let $\Delta \phi_H$ be the variation in the service quality due to variations in the number of requests served in time, and $\Delta \phi_L$ be the variation in the service quality due to variations in the number of requests served late. Then, the overall variation in the service quality is as follows.

$$\Delta \phi = \frac{\partial \phi}{\partial H} \Delta \phi_H + \frac{\partial \phi}{\partial L} \Delta \phi_L. \quad (15)$$

In order to find the maximum and/or minimum values of the service quality, the derivatives have to be equal to 0. Using (11), we have,

$$\frac{\partial \phi}{\partial H} = 0.$$

Or,

$$\frac{L}{(H+L)^2} = 0.$$

This gives rise to $L = 0$. The second derivative of service quality for $L = 0$ comes out to be $+ve$, signifying that at $L = 0$, there is a maxima for the service quality. Putting $L = 0$ in (9), we get,

$$\begin{aligned}
\phi_{\max} &= \frac{cH}{(c+p)H} \\
&= \frac{c}{c+p}. \quad (16)
\end{aligned}$$

For $c = p$, we have, $\phi_{\max} = \frac{1}{2}$. In other words, when the number of credits per request served in time is equal to the number penalty points per request served late, then the maximum value of the service quality is $\frac{1}{2}$.

Similarly, equating the partial derivative of service quality given by (12) to 0, we get $H = 0$. At $H = 0$, the second derivative of the service quality is $-ve$, therefore, there is a minima for the service quality at $H = 0$. Putting $H = 0$ in (12), we get the minimum value of the service quality which is as follows.

$$\begin{aligned}
\phi_{\min} &= -\frac{pL}{(c+p)L} \\
&= -\frac{p}{c+p}. \quad (17)
\end{aligned}$$

For $c = p$, we have, $\phi_{\min} = -\frac{1}{2}$. The minimum and maximum values of service quality, as given by (16) and (17), confirm our earlier argument that $-\frac{1}{2} \leq \phi \leq \frac{1}{2}$ for credit-based service quality model.

In what follows, we present results and discussion.

## IV. RESULTS AND DISCUSSION

The eLibrary gathers information about the service profiles of different types of contents provided by the eLibrary. Based on the information gathered for a certain period of the observation time, statistical analysis of the service profiles is performed so as to improve the service quality provided by the eLibrary to its users. The information gathered in a manner described above is analyzed. The library keeps track of how many hits were made by users with different profiles and how many requests were timely satisfied and for how many requests arrangements/procurements from else where were made and how many requests were not satisfied at all. What type of requests were the most frequent and what type of requests were less frequent. The analysis of service profiles of the contents requested by the users and provided by the eLibrary is carried out in order to evaluate the service quality provided by the eLibrary to its users. Based on the average value of the service quality, measures can be adopted to improve the service quality of the eLibrary.

Let us examine the how the service quality varies in the delay based model. Figure 1 shows the service quality as a function of the number of days by which the service of requests is delayed, where the number of penalty points for each request is one per day or two per day. We observe that as the delay in the number of days is increased, the service quality decreases exponentially. Also, we observe that service quality decreases more rapidly if the number of penalty points is increased from one penalty point per day to two penalty points per day. Note that when the number of penalty points per day is 1, the maximum value of the service quality is 2, and when the number of penalty points per day is $\phi_{\max} = 2$, the maximum value of the service quality is $\phi_{\max} = 3$.

Figure 2 shows the service quality as a function of the number of penalty points for each request served late, where the requests are delayed by one day or two days. We observe that the service quality is 1 for the number of penalty points equal to 0, and after that it reaches to its normal value at penalty points equal to 1. After that, the service quality starts decreasing with the number of penalty points for a given number of days by which the request is late.

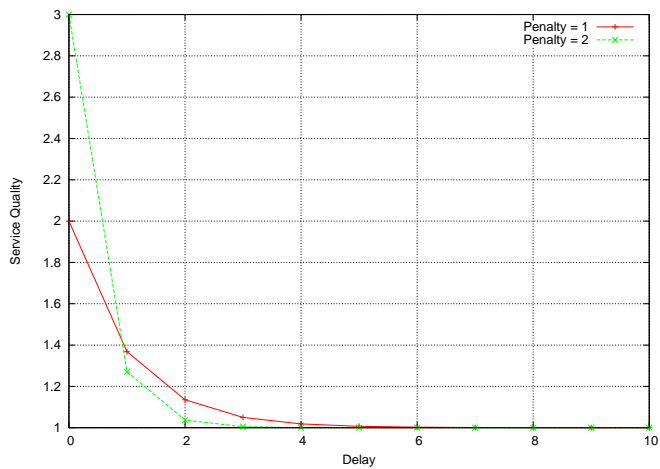

Fig. 1. Service quality as a function of the number of days by which the service of requests is delayed, where the number of penalty points for each request is one per day or two per day (*Delay Based Model*).

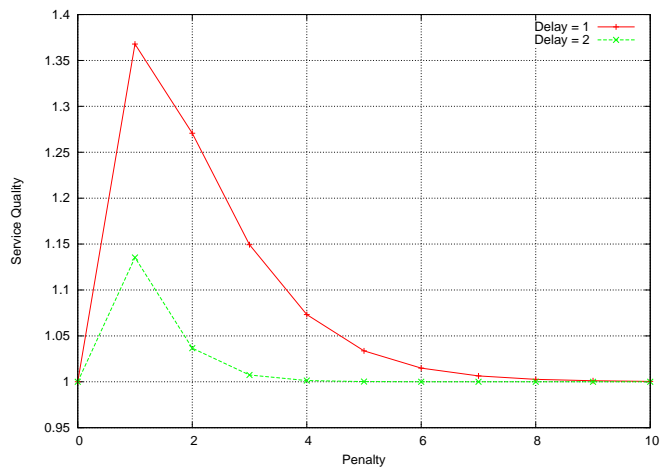

Fig. 2. Service quality as a function of the number of penalty points for each request served late, where the requests are delayed by one day and two days (*Delay Based Model*).

However, the delay based service quality model uses only one parameter, namely, penalty points, and more is the number of penalty points per day, the maximum value of the service quality is set higher. This seems reasonable in the sense that if the decrement for the service quality (which is the number of penalty points per day), the maximum value from which the service quality should start decreasing, is set higher as compared to the situation where the decrement in the service quality is relatively small as there is no way to increase the service quality. In other words, the larger value of the number of penalty points per day also plays the role of implicit credits: higher the penalty, larger is the value of the maximum service quality. There are no explicit credits. This is analogous to a bankers cash: larger the rate of withdrawal from the bank, more cash the banker should have with himself/herself to start with. It is possible that a banker giving away money to his/her customers at a higher rate may finish his/her start money more rapidly as compared to the one from whom rate of withdrawal is smaller and who starts with a smaller money.

We now examine the service quality in the credit based model, where the library is assigned explicit credits when the requests are served in a timely fashion, and explicit penalty points when the item to be served by the library is late. Figure 3 shows the service quality of an eLibrary as a function of the number of requests that were served late for the credit based model. The number of requests served in a timely manner is taken to be 10 and 20. We observe that as the number of requests served late is increased, the service quality the eLibrary decreases. At a certain point in time, the service quality becomes negative. It means that the service quality has been deteriorated significantly and corrective measures should be taken to improve the service quality of the eLibrary. In what follows, we conclude the paper.

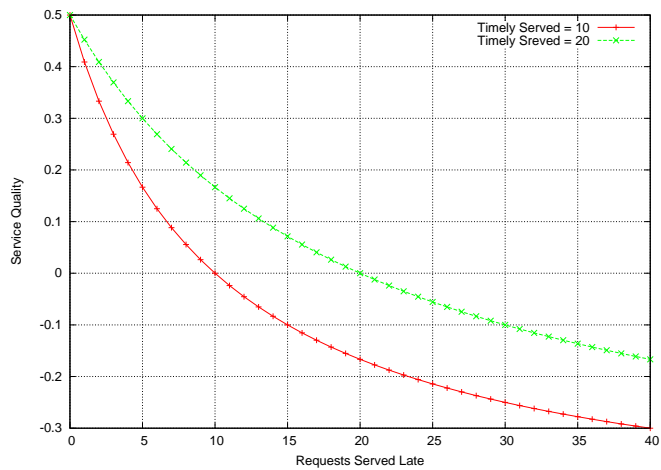

Fig. 3. Service quality as a function of the number of requests served late, where the number of requests served in a timely manner is 10 and 20 (*Credit Based Model*).

## V. CONCLUSION

Devising a scheme for evaluating the service quality of an institutional electronic library is a difficult and challenging task. The challenge comes from the fact that the services provided by an institutional electronic library depend upon the contents requested by the users and the contents housed by the library. Different types of users might be interested in different types of contents. In this paper, we propose a technique for evaluating the service quality of an institutional electronic library. Our scheme is based on the service profiles of contents requested by the users at the server side (i.e. service profiles are maintained at the server of the eLibrary and not at the side of the end user). For analyzing the service quality, we presented two analytical models. The first one is based on the number of days by which the item to be served by the library is delayed and the penalty points per day. The second model is based on the credits earned by the library if the item is served in a timely fashion, and the penalties if the item is delayed. These models may help in evaluating the service quality of the eLibrary and taking the corrective measures to improve it.